# Impacts of dielectric screening on the luminescence of monolayer WSe$_2$

Fábio J. R. Costa[1], Thiago G-L. Brito[1], Ingrid D. Barcelos[2] and Luiz Fernando Zagonel[1]

[1]Gleb Wataghin Institute of Physics - University of Campinas – UNICAMP, Campinas 13083-859, Brazil

[2]Brazilian Synchrotron Light Laboratory (LNLS), Brazilian Center for Research in Energy and Materials (CNPEM), 13083-970 Campinas, SP, Brazil

*Abstract—* Single layers of transition metal dichalcogenides, such as WSe$_2$ have gathered increasing attention due to their intense electron-hole interactions, being considered promising candidates for developing novel optical applications. Within the few-layer regime, these systems become highly sensitive to the surrounding environment, enabling the possibility of using a proper substrate to tune desired aspects of these atomically-thin semiconductors. In this scenario, the dielectric environment provided by the substrates exerts significant influence on electronic and optical properties of these layered materials, affecting the electronic band-gap and the exciton binding energy. However, the corresponding effect on the luminescence of transition metal dichalcogenides is still under discussion. To elucidate these impacts, we used a broad set of materials as substrates for single-layers of WSe$_2$, enabling the observation of these effects over a wide range of electrical permittivities. Our results demonstrate that an increasing permittivity induces a systematic red-shift of the optical band-gap of WSe$_2$, intrinsically related to a considerable reduction of the luminescence intensity. Moreover, we annealed the samples to ensure a tight coupling between WSe$_2$ and its substrates, reducing the effect of undesired adsorbates trapped in the interface. Ultimately, our findings reveal how critical the annealing temperature can be, indicating that above a certain threshold, the heating treatment can induce adverse impacts on the luminescence. Furthermore, our conclusions highlight the influence the dielectric properties of the substrate have on the luminescence of WSe$_2$, showing that a low electrical permittivity favours preserving the native properties of the adjacent monolayer.

## I. INTRODUCTION

Over the past decade, the research focused on van der Waals (vdW) materials has grown into a very active field, embracing studies on this continuously growing class of materials[1]–[5]. Most of these systems exhibit a wide range of attributes that are thickness dependent, and this degree of freedom allows the use of the number of layers to tune the properties of such vdW's materials. A typical case is that of semiconducting transition metal dichalcogenides (TMDCs), such as MoS$_2$ and WSe$_2$. These materials exhibit a transition from an indirect to a direct electronic band-gap[6]–[8], that takes place when the crystal is thinned down from bulk to a monolayer (ML), leading to intense light-matter interactions[6, 9]–[14], making them promising candidates for light emitting devices. The reduced thickness of these systems makes them highly sensitive to the surrounding environment[15]–[22]. Therefore, the choice of an appropriate substrate for the deposition/growth or mechanical transfer of the TMDC-MLs can have critical effects on the properties of the assembled heterostructure. These impacts are strongly related to the interaction happening on the interface between the two materials, affecting the exciton dynamics within the layered semiconductor, and consequently its light-emitting capabilities. Consequently, the choice of an appropriate substrate can be used to tune key properties of the system.

This susceptibility allows manipulating properties of the TMDC by changing the surrounding dielectric environment. There is currently a consensus that an increasing dielectric screening leads to a systematic reduction of the electronic band-gap ($E_g$), accompanied by a decrease of similar magnitude in the exciton binding energy ($E_b$)[4, 23]–[32]. However, the impacts on the optical band-gap ($E_{opt}$) are still under debate, as several studies account for a red-shift in response to a rising dielectric environment[4, 26]–[28, 30, 33], while others report an opposite effect[23, 24, 31, 34], and in some cases, even the lack of a clear correlation is observed[21, 24, 28, 32, 35, 36]. While several external effects are known to induce shifts in the optical band-gaps in TMDCs, the coupling with the substrate should not be overlooked. Concerning the screening effect induced by the underlying material, it has been theoretically predicted[27, 29] that this influence is highly sensitive to the distance between the layered semiconductor and its underlying support. The experiment performed by Borghardt *et al*[28] supports this interpretation. Their study reports on a consistent red-shift of the excitonic resonances in response to increasingly permitting hydrophobic substrates. When using hydrophilic supports, no clear correlation was observed. This observation highlights that the trapping of adsorbates in the interface can affect the way the charges within the TMDC perceive the dielectric properties of the adjacent media.

Indeed, the observation of residues trapped in the interface of vdW's materials and surfaces has been widely reported[37]–[46]. The general conclusion is that these adsorbates can compromise the efficiency of the assembly, introducing extrinsic disorder[43] and acting as a decoupling media between the layers composing the heterostructure. The studies performed by Tongay *et al*[37] and Román *et al*[45] show that thermal annealing can be an effective strategy to remove these contaminants, ensuring a tight interlayer coupling in heterostructures containing TMDC-MLs. However, the conditions where this treatment is applied are diverse within the literature: reports of the temperature used to perform annealing on samples containing

TMDCs vary within a wide range[21, 33, 37, 44, 45, 47]–[49]. Even though there is a well-established comprehension that this strategy is a successful way of driving out undesirable contaminants from the interfaces on heterostructures, a comprehensive study focused on the impacts of the heating on the TMDC itself is missing.

The present work aims to study the impacts of different underlying materials on the luminescence of ML-WSe$_2$. Several materials were used as substrates, providing a wide range of electrical permittivities, and all samples were annealed, to ensure an effective coupling between TMCD and its supports. The impacts of the thermal treatment were examined by performing several annealing cycles at increasing temperatures. This analysis showed that the luminescence can be severely suppressed above a certain threshold temperature. Moreover, our data show that the dielectric environment affected not only the energy of the excitonic emission but also its intensity. Ultimately, our results contribute to deepen the understanding of how the dielectric surrounding can impact the luminescence of MLs of TMDCs.

## II. METHODS

**Sample preparation**: Flakes of tungsten disselenide (WSe$_2$) were mechanically exfoliated from bulk crystals from HQ Graphene. These flakes were then transferred to a Polydimethylsiloxane (PDMS) stamp, where optical microscopy was employed to locate and identify MLs by their contrast. Afterwards, the selected MLs were deposited on specific positions of the substrates. The current study comprised a wide set of supports, ranging from insulating materials such as silicon dioxide (SiO$_2$), hexagonal boron-nitride (h-BN) and Mg$_3$Si$_4$O10(OH)$_2$ (Talc), to conducting substrates, such as gold (Au), highly oriented pyrolytic graphite (HOPG) and indium tin oxide (ITO).

**Thermal annealing**: The annealing of the samples was performed in a custom-built oven. The samples were inserted within a vacuum chamber, and after the pressure reached $10^{-7}$ torr, the annealing was initiated. Four heating cycles were performed, where the samples were kept in the thermal treatment for 12 hours at each step, under the following temperatures: 400 K, 430 K, 460 K and 490 K.

**Photoluminescence spectrocopy**: The PL measurements were carried out in ambient conditions, using a custom micro-PL setup. The excitation source was a 2.33 eV (532 nm) continuous wave laser diode. The power employed in the experiments was kept close to 200 $\mu$W. The PL signals were recorded by an Andor Newton 970 EMCCD, mounted to an Horiba iHR 320 spectrometer. The spectra acquired were calibrated in energy, using the spectral lines of an Ocean Insight HG-2 Mercury Calibration Source as a reference. The procedure for such correction is described elsewhere[50, 51]. The reproducibility of the PL intensities were monitored before each acquisition by performing control measurements on a bulk Al$_2$O$_3$ (sapphire) sample.

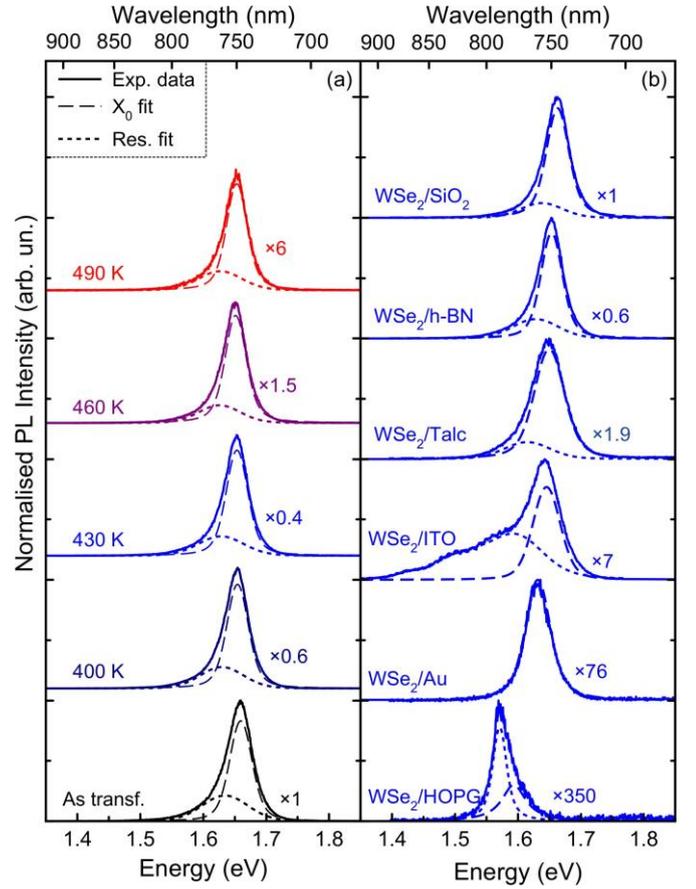

Fig. 1. PL spectroscopy of WSe$_2$ monolayers. (a) Series of spectra acquired for ML-WSe$_2$/h-BN as a function of different annealing temperatures. (b) Series of spectra acquired on ML-WSe$_2$ samples annealed at 430 K deposited onto several substrates. All measurements were performed at ambient conditions. The residual fitting includes trions and other contributions observed in energies below the main $X_0$ peak.

## III. RESULTS AND DISCUSSION

A set of representative results acquired within this study is summarised in Fig. 1. Fig. 1 (a) exhibits a set of PL measurements performed in a ML-WSe$_2$ deposited on top of h-BN, as a function of successive annealing temperatures. The contributions due to neutral ($X_0$) and charged excitons ($T$), alongside other species, were assigned by adjusting the data as a convolution of individual peaks[21, 33, 44, 45, 48, 52, 53]. In the present analysis, Voigt functions were used to model each of these contributions to the overall PL spectra. Concerning the shape of the PL spectra, the comparison presented indicates a great similarity of the data acquired after successive annealing steps. While this was a common tendency between the samples transferred on insulating supports, a distinct behaviour was observed for the monolayers deposited on conducting substrates. The spectra acquired for the complete set of samples are presented in Fig. S1, in the supplementary material, along with brief comments on these effects.

Fig. 1(a) presents the impact of the annealing temperature on the PL intensity of a ML-WSe$_2$ transferred onto h-BN.

The data reveals that the first two annealing steps led to a significant increase in the PL intensity, which can be attributed to the removal of contaminants that somehow compromised the quantum efficiency of the TMDC. The successive heating of the sample at higher temperatures led to a severe decrease in the PL intensity. This latter tendency was common among all the samples studied, regardless of the substrate, implying that such commonly used temperatures can exert unfavourable impacts on the TMDC quantum efficiency.

Fig. 1 (b) presents a comparison of the PL spectra acquired for different samples, as a function of the substrate. These measurements were performed after the annealing at 430 K. Such temperature is enough to reduce the influences of a decoupling layer of contaminants between TMDC and support, being also below the threshold where adverse impacts rose. This data highlights how the influence of underlying materials of various natures can affect different aspects of the interaction between excitons and holes within the TMDC, leading to several impacts observed in the luminescence of these systems. Additionally, measurements were performed as a function of the excitation power (Fig. S4), but no evident variation on the spectra was observed.

The analysis of quantitative aspects of these spectra highlighted that the surrounding environment exerts considerable influence on the doping state of the TMDC[52]. For instance, the suppression of emissions due to trions was observed for the $WSe_2$/Au sample. On the other hand, a dominant contribution due to transitions below the $X_0$ energy emerged from MLs-$WSe_2$ deposited on ITO and HOPG. Considerable influences of the substrate were also evident in the energy and intensity of these transitions. The following analysis indicates a strong relationship between these effects and the dielectric properties of the underlying materials.

The following sections will cover the lines of investigation presented in the introduction. Initially, a wide range of substrates was employed to study the impacts of several dielectric environments on the luminescence of MLs of $WSe_2$. Such analysis was performed on annealed samples, ensuring a reduced influence of possible contaminants present in the interface between $WSe_2$ and its underlying materials. Secondly, the impacts of the thermal treatment itself were investigated. Samples were exposed to a set of annealing cycles at increasing temperatures, providing insights into the implications of the treatment conditions on the luminescent properties of the TMDC.

*A. Environmental screening impacts*

To elucidate the impacts of diverse dielectric surroundings on the luminescence of MLs-$WSe_2$, a set of annealed samples was examined by PL spectroscopy. These measurements are presented in Fig. S1, and its analysis is summarised in Fig. 2, highlighting relevant effects both on energy and intensity. These quantities were plotted against the high-frequency dielectric constants of $SiO_2$[25, 54], h-BN[25, 55, 56], ITO[57, 58], Au[59] and HOPG[60], and a systematic red-shift of $E_x$ is evidenced in Fig. 2 (a). Here we considered a close interaction between MLs and their underlying supports, neglecting possible impacts of local variations due to the roughness of the substrates. Therefore, assuming a perfect adhesion between the layers of the heterostructure, the dielectric constants used represent an upper limit to the electrical permittivity felt by the MLs-$WSe_2$.

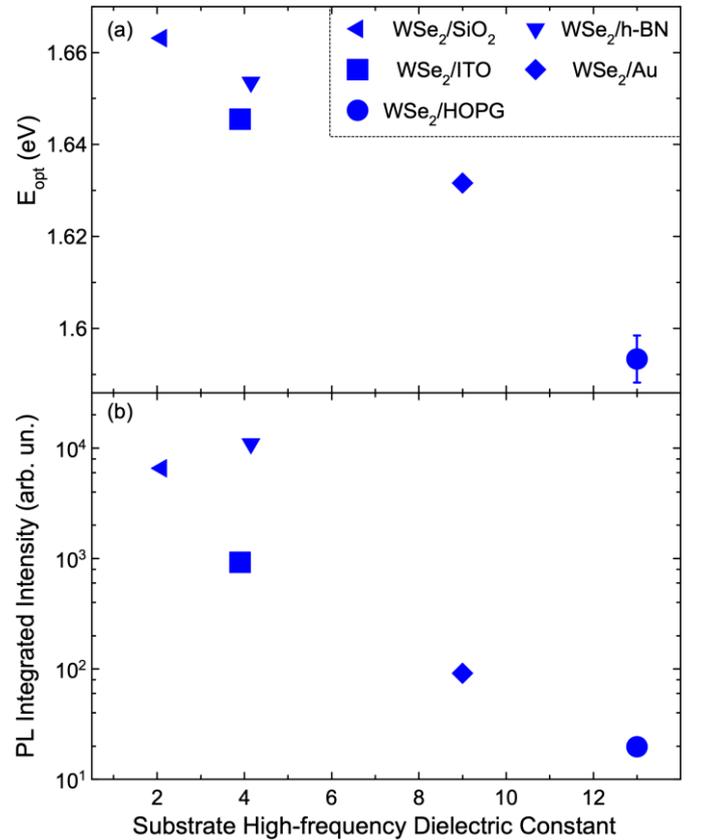

Fig. 2. Influence of the dielectric environment on the luminescence of MLs of $WSe_2$ annealed at 430 K. (a) Optical band-gap and (b) Integrated PL intensity as functions of the substrate's high-frequency dielectric constant. In most cases, the uncertainties in energy and intensity were negligible and were omitted.

This displacement of the excitonic peak to lower energies is strongly related to the permittivity of the environment surrounding the MLs-$WSe_2$, as these conditions affect the interaction of electrons and holes within the TMDC. On one hand, the increasing screening attenuates the Coulomb interaction between these charges, leading to weaker bonds, and ultimately reducing the binding energy ($E_b$) of the quasiparticles[4, 26]. On the other hand, the increased screening also affects electron-electron interactions, leading to a renormalization of the electronic band gap ($E_g$), with decreasing magnitude[23, 30].

It is noteworthy that these changes are of similar magnitude, and due to the relationship between $E_g$, $E_b$ and $E_{opt}$

$$E_{opt} = E_g - E_b \qquad (1)$$

this difference in energy defines the influence the screening has on the optical band-gap. Our data exhibits a tendency where the excitonic peak of WSe$_2$ shifts towards lower energies in response to an increasing permittivity of the surroundings. Ultimately, this trend implies that in the current scenario the electronic band-gap was more sensitive to the screening than the exciton binding energy.

This analysis also comprised the impacts of the dielectric surrounding on the quantum efficiency of each heterostructure, assessed through the integrated PL intensity. The tendency presented in Fig. 2(b) indicates a correlation between these two quantities, implying that a higher dielectric screening contributes to a systematic shrinkage of the PL intensity, covering a couple of orders of magnitude.

A central aspect of this interpretation is related to the choice of the dielectric constants for the materials used as supports. The current literature on this topic presents controversies regarding the scale of frequencies where the screening of electron-hole interactions takes place: while several publications consider the static limit, others employ the dielectric constants at the high-frequency regime. The work published by Lin *et al*[24] illustrates how critical this choice can be. In this publication, the influence of the surrounding's permittivity on $E_{opt}$ changes depending on the set of dielectric constants included in the analysis.

This inconsistency is frequently found across the literature, where several publications adopt the static values[24, 28, 31] for the dielectric constants, while several others employ the high-frequency values[24, 25, 32, 61]. The latter interpretation is supported by the argument that the dielectric response of the medium to the exciton will occur in a frequency range close to $E_b$[62]. Considering WSe$_2$ as an example, its exciton binding energy is ∼500 meV (∼THz). At this regime, the lattice polarisation is not fast enough to interfere with the electron-hole pair, and the contribution due to the electronic motion predominantly affects this dynamic[62]. For completeness, Fig. S2 in the Supplementary Information presents the data used in Fig. 2, but this time employing static values of the dielectric constants. The observed discrepancy evidences how the choice between static and high-frequency dielectric constants can influence the current analysis.

Such discrepancies highlight the necessity of reaching a consensus concerning the adequate frequency range to be considered in this scenario. To surpass this limitation, Fig. 3 was produced, taking advantage of the fact that both the optical band-gaps and the integrated PL intensities are quantities precisely extracted from the experimental data. Therefore, the relationship presented in this plot offers conclusions less dependent on external factors, such as the choice of proper values for the dielectric constants. As a first conclusion, this data evidences a clear relationship, where the samples presenting the highest excitonic transition energies were also those producing the most intense PL emission.

To further interpret this relationship, it is noteworthy that the magnitude of the dependence of the optical band gap concerning changes in the surrounding dielectric environment is still under discussion. Several publications report shifts of dozens of meVs for a wide range of permitivitties[4, 17, 26, 28, 31], while others could not establish a direct correlation between both effects[21, 24, 32, 35, 36]. Furthermore, the nature of this impact is also a topic under debate. While several reports account for a rise in energy of the optical band gap in response to an increasing dielectric surrounding[23, 24, 31, 34], others mention the opposite effect[4, 26]–[28, 30, 33]. Our observations agree with the latter interpretation, as the increasing screening provided by the supports led to a systematic red-shift of the optical band gap.

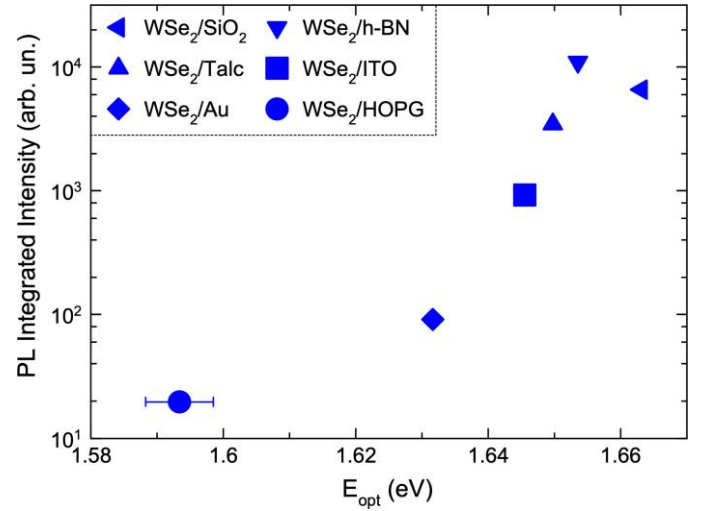

Fig. 3. Relationship between the optical band-gap and the integrated PL intensity acquired for MLs of WSe$_2$ annealed at 430 K for a range of substrates. In most cases, the uncertainties in energy and intensity were negligible and were omitted.

In light of this interpretation, the data presented in Fig. 3 can be analysed as follows: under the influence of a more intense screening, $E_{opt}$ experiences a red-shift, providing an indirect measurement of the screening acting in the system. This relationship has been demonstrated in several calculations[4, 30] and experiments[26, 28], and also at Fig. 2(a). Therefore, given the interdependence between the exciton binding energy and the electronic and optical band-gaps (Eq. 1), the small red-shifts observed for the latter can be directly related to wider reductions being experienced by both $E_g$ and $E_b$[4, 27, 30].

Consequently, the energies measured for $E_{opt}$ can be related to the screening felt by the TMDC and to a further reduction of $E_b$. This suggests a direct relationship between $E_b$ and the PL intensity: the samples whose optical band-gaps were more red-shifted (and whose exciton binding energies were more reduced) due to the screening of the substrate exhibited the weakest PL intensity. Therefore, this data implies that reducing $E_b$ leads to a drastic reduction of the PL intensity.

Ultimately, this interpretation highlights the contribution of the dielectric properties of the substrate to the optical efficiency of the adjacent TMDC. Concerning the role of the underlying material in preserving the native properties of the TMDC, aspects such as the absence of dangling bonds, low-roughness and inertness are desired. As h-BN reunites all of these[30, 33, 63]–[65], it has been widely used as a substrate[1, 4]. Recently it has been shown that Talc[5, 66, 67], a naturally occurring wide band-gap layered silicate, also possesses such features, being considered a viable low-cost alternative to h-BN as a substrate[68]–[70].

Thus, a direct comparison of the performance of $WSe_2$ deposited onto these similar materials can offer insights into the impact of the dielectric properties of these efficient substrates. Fig. 3 shows that a shift of $\sim 4$ meV is present between these two scenarios, implying that Talc exerts a more intense screening than h-BN. It would also implicate that the silicate material induces a greater reduction of the exciton binding energy of the neighbouring $WSe_2$-ML. Moreover, this decrease in $E_b$ would also be related to the $\sim 70\%$ less intense PL emission measured for the $WSe_2$/Talc sample. In a broader picture, this result implies that alongside the aspects shared between these two wide band-gap layered substrates, providing a lower dielectric screening is key to preserving the PL of the adjacent TMDC.

To conclude, we emphasize that the impact of dielectric screening is regarded as the main reason for the observed shift in the excitonic transition energy. It is noteworthy that several other factors can tune the light emission properties of ML-TMDCs, such as charge doping and strain. In the first case, the balance of negative/positive charges within the TMDC can be successfully controlled through electrostatic gating[71], allowing the manipulation of the light emission properties. Such experiments showed that for a wide range of induced doping levels, the resultant shift of the excitonic peak reached a couple of dozens of meVs[72]–[76], which is significantly smaller than the displacements observed here.

Another relevant way to tune the optical band-gaps in TMDCs is via strain engineering, which leads to significant reductions in the electronic band gap, with a lesser impact on the exciton binding energy[4, 71, 77]. Consequently, this effect could also be considered a source for the shifts observed in our data. For the case of $WSe_2$, the displacement of the exciton resonance per percent of strain applied lies around 47-54 meV/%[78, 79]. Some publications[15, 80] report that PDMS transfer of TMDCs usually imposes a residual tension on the layered semiconductors, with values ranging a strain of 0.2%. Considering this value as a reasonable estimation for the strain acting on the samples investigated within this study, one could expect the PL peaks to be shifted by 10 meV. Therefore, considering the typical impacts from doping and strain, the shifts in the excitonic transition energy observed here ($\sim 60$ meV) could not result solely from either of these sources.

*B. Annealing impacts*

The impacts of exposing the samples to successive annealing cycles at increasing temperatures are summarised in Fig. 4. Fig. S3 in the Supplementary Information presents the complete set of spectra. The main impacts observed can be analysed in two distinct stages: initially, the first annealing cycles affected the PL intensities in various ways, giving rise to distinct tendencies. The subsequent annealing cycles lead to a systematic drop in the PL intensities for all samples.

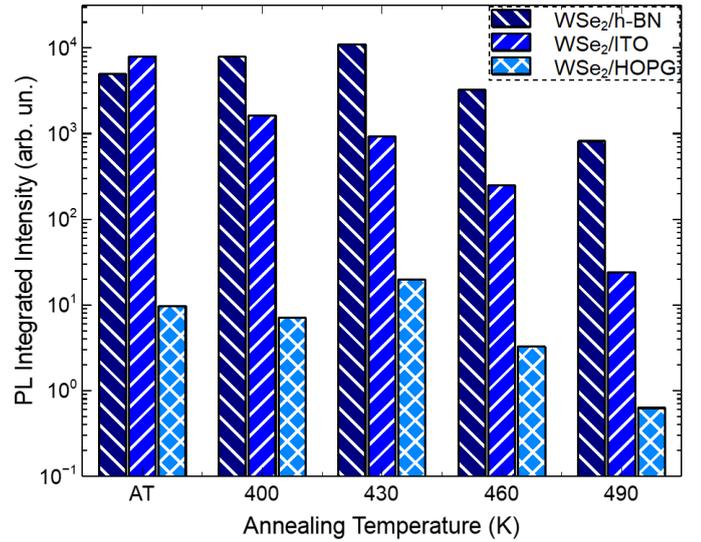

Fig. 4. Evolution of the PL integrated intensity of MLs of $WSe_2$ for several supports, as a function of annealing temperatures.

Based on these observations, the impact of the increasing treatment temperatures can be interpreted as follows: the first annealing cycles acted as a cleaning strategy, removing adsorbates from the interface, and producing a tighter coupling between TMDC and substrate. This process improved the sample homogeneity, favouring the luminescence arising from $WSe_2$[33], as observed for the $WSe_2$/h-BN sample. Concerning the conducting substrates, alongside the reduction of the exciton binding energy due to the increased screening, other effects can also result in the observed quenching of the luminescence. For instance, the coupling with noble metals such as Au is known to give rise to interactions between the electronic states of the semiconductor and the underlying material[81]. This mechanism allows the opening of non-radiative recombination channels between TMDC and the substrate, contributing to an increase in non-radiative recombination processes, consequently weakening the luminescence[82]–[84].

Finally, the ML transferred to HOPG displayed a somewhat peculiar behaviour, exhibiting the weakest PL among all samples investigated, with intensities a couple of orders of magnitude lower than the other samples. This effect could be related to the hydrophobic character of graphite, which disfavors the imprisonment of polar adsorbates between the TMDC and HOPG. Considering this scenario, the $WSe_2$-ML could be in closer contact with the substrate even before the first annealing. Then, the first two annealing cycles would have reduced impacts, as observed in the experiment.

The further increase in the annealing temperatures produced a systematic reduction of the PL intensities in all samples. A possible cause for this suppression could be the creation of a high density of defects. Several publications have reported that the exposure of TMDCs to heating treatments could produce this outcome[44, 85]–[89]. However, alongside this effect, an increasing density of defects in a TMDC usually leads to a doping effect[23, 44, 86, 90]–[93]. A typical effect of varying the concentration of charge carriers in a TMDC is to drastically change the proportion of X/T emissions in the PL spectrum[44, 52, 71, 92].

Therefore, assuming that the annealing cycles performed above the 430 K temperature could lead to the formation of defects in the MLs-WSe$_2$, a significant variation in the X/T populations would be expected. However, a common trend was not observed here, as reported in supplementary figures S1 and S2, where several distinct outcomes were identified: some samples exhibited relative stability of the PL composition, while others exhibited meaningful variations of either X/T contributions. Hence, the diversity of outcomes observed prevents the determination of a clear link between the formation of defects and the fluctuations observed in the PL intensities after each annealing step.

## IV. Conclusions

This study comprises a broad investigation of the luminescence of heterostructures containing MLs of WSe$_2$, including a wide range of different supports. The excitonic transition energies measured from these various heterostructures provided clear evidence of a systematic red-shift of the optical band-gap as the environmental screening increased. This result implies that the surrounding dielectric environment exerted a higher influence on the electronic band-gap than on the exciton binding energy. Moreover, the relationship between the optical band-gap and the PL intensity exhibited a clear tendency, implying that increases in the dielectric environment also contribute to the suppression of the luminescence. This effect was interpreted as a manifestation of the reduction of the exciton binding energy, indicating that as the electron-hole interactions become weaker, the PL yield also decreases. Ultimately, this result suggests that low permittivity is a desired property for substrate materials for TMDCs. Regarding the impacts of the annealing on the luminescence of the WSe$_2$ heterostructures, a careful evaluation of the changes in the PL profiles after each heating cycle provided important insights into the effects of this treatment. Additionally, samples heated above 430 K exhibited significant suppression of the PL intensity. Even though the origin of this effect remained elusive, our data indicate that a careful choice of temperature is crucial to preserve the light-emitting capabilities of TMDC-based heterostructures during annealing.

## V. Data availability statement

The data that support the findings of this study are openly available at the following URL/DOI: https://doi.org/10.25824/redu/ZRVPKG.

## VI. Acknowledgements

This work was supported by the Fundação de Amparo à Pesquisa do Estado de São Paulo (FAPESP) grant 2021/06893-3 and by the Coordenação de Aperfeiçoamento de Pessoal de Nível Superior (CAPES), grant 88887.517233/2020-00. The authors also would like to acknowledge the Brazilian Synchrotron Light Laboratory (LNLS) for the LAM facilities in experiments involving 2D sample preparations. I. D. B. acknowledges the financial support from the Brazilian Nanocarbon Institute of Science and Technology (INCT/Nanocarbono), FAPESP (grant 2019/14017-9), CNPq (grant 311327/2020-6), and the award of L'ORÉAL-UNESCO-ABC for Women in Science - Brazil/2021.

# Supplementary Material:

# Impacts of dielectric screening on the luminescence of monolayer WSe2

Fábio J. R. Costa[1], Thiago G-L. Brito[1], Ingrid D. Barcelos[2] and Luiz Fernando Zagonel[1]*

[1]*Applied Physics Department, Gleb Wataghin Institute of Physics, University of Campinas—UNICAMP, 13083-859 Campinas, São Paulo, Brazil.*
[2]*Brazilian Synchrotron Light Laboratory (LNLS), Brazilian Center for Research in Energy and Materials(CNPEM), 13083-970 Campinas, SP, Brazil*

E-mail: fabiojrc@ifi.unicamp.br, zagonel@ifi.unicamp.br

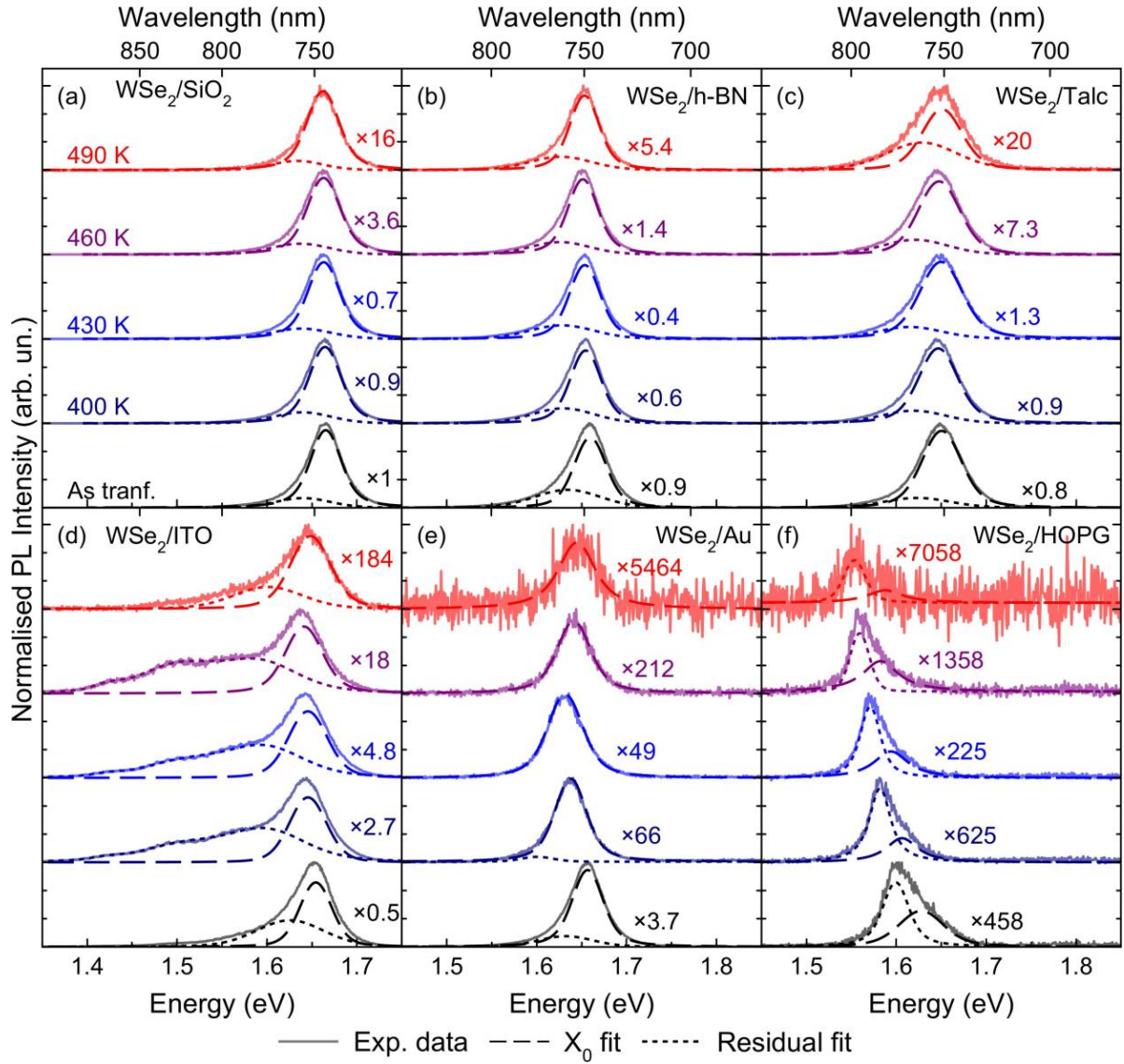

Figure S1: Evolution of the PL of ML-WSe$_2$ samples as a function of annealing temperatures. Intensities are normalised with respect to the PL of the as-transferred ML-WSe$_2$/SiO$_2$ sample. All data was taken at ambient conditions with the same laser power.

Fig. S1 summarises the PL measurements performed in the complete set of samples, highlighting the impacts on the luminescence arising both from the different supports and from the increasing annealing temperatures. Most of the spectra were fitted as a combination of two Voigt peaks, representing the contributions of neutral and charged excitons. The exception was the data acquired on the ML-WSe$_2$/ITO sample, which exhibited a more complex emission, requiring more peaks to achieve an appropriate fitting. As the precise identification of these contributions was beyond the scope of the present study, they were simply referred to as residual contributions.

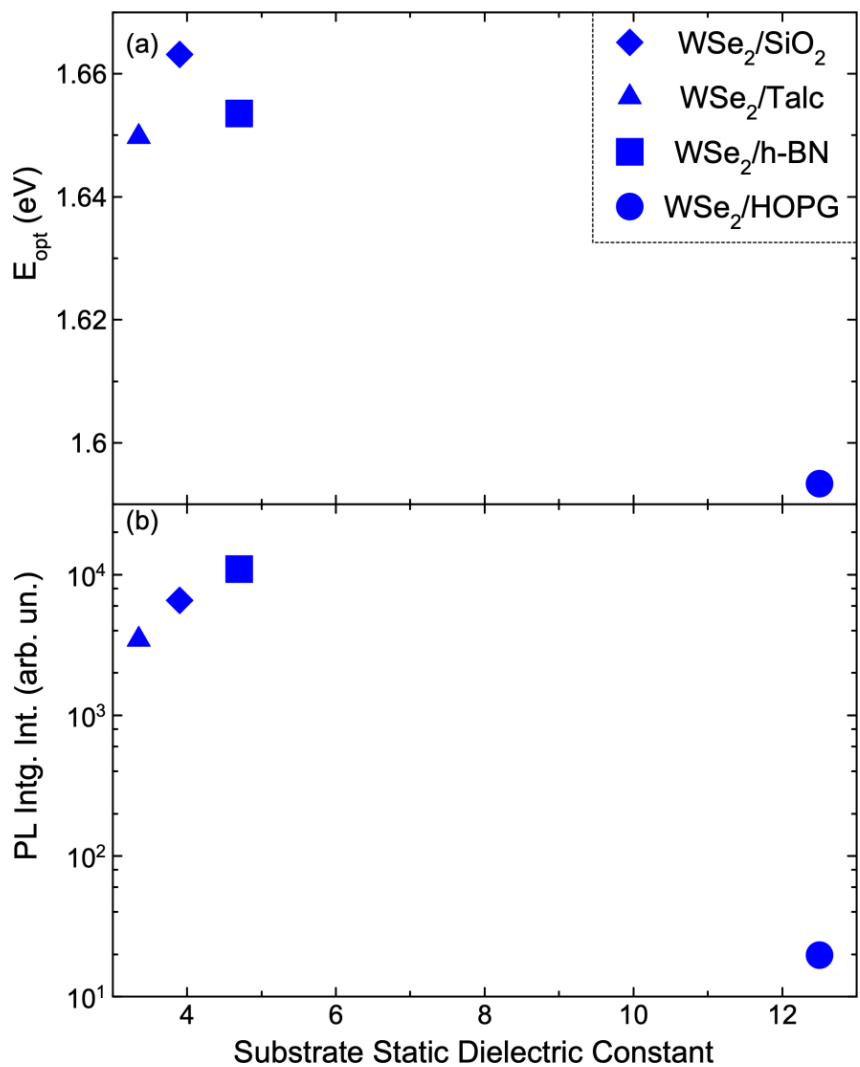

Figure S2: Influence of the dielectric environment on the optical properties of ML-WSe$_2$. (a) Optical band-gap and (b) Integrated PL intensity as functions of the substrate's DC dielectric constant.

Fig. S2 displays the influence of the dielectric environment on the luminescence of MLs-WSe$_2$. This analysis considers the static values for the dielectric constants of the supports. This choice led to a contrasting behaviour when compared to the one presented in Fig. 2 in the main text. Considering Fig. S2 (a), the impact of the increasing dielectric environments on the energy of the excitonic transition was less clear than the one observed in the main text: while for the samples transferred onto insulating supports (SiO$_2$, h-BN and Talc) a slight blue-shift was observed, the sample deposited onto HOPG suffered a massive red-shift.

This apparent contradiction was also evident on Fig. S2 (b), where the PL integrated intensity rose as the dielectric constant of the insulating supports was increased. Nevertheless, the sample deposited onto HOPG displayed a contrasting behaviour, exhibiting the weakest PL intensity. These conflicting outcomes between the analysis performed with the static or high-frequency values for the dielectric constants highlight how critical this choice can be.

Fig. S3 summarises the impacts of the annealing steps on the luminescence of heterostructures containing MLs-WSe$_2$, including the complete set of samples studied here. Fig. S3 (a) displays the evolution of the PL intensities for each annealing step, grouped by temperature. Fig. S3 (b) exhibits the neutral exciton contribution to the overall spectra, where several distinct behaviours were observed. The main feature highlighted by this analysis is the lack of a clear correlation between the PL intensity and the $X_0$ contribution to the spectra, as discussed in the main text.

For instance, considering the behaviour exhibited by ML-WSe$_2$ deposited on SiO$_2$ and Talc. Focusing on the initial annealing steps, up to the temperature of 430 K, both samples exhibited contrasting behaviours regarding the PL intensity: while the first experienced an increase, the second exhibited a falling tendency. This contrast was not reproduced by the $X_0$ contribution to the overall PL spectra, where both samples exhibited relative stability along these steps. Considering the whole picture presented by this data set, the overall trend observed is that the fluctuations in the PL intensity and $X_0$ ratio are not directly correlated.

Finally, Fig. S3 (c-h) summarises the impacts of the subsequent annealing cycles on the PL intensity for each sample studied. Grouping this data for each sample separately highlights the influence of different substrates during the annealing cycles conducted.

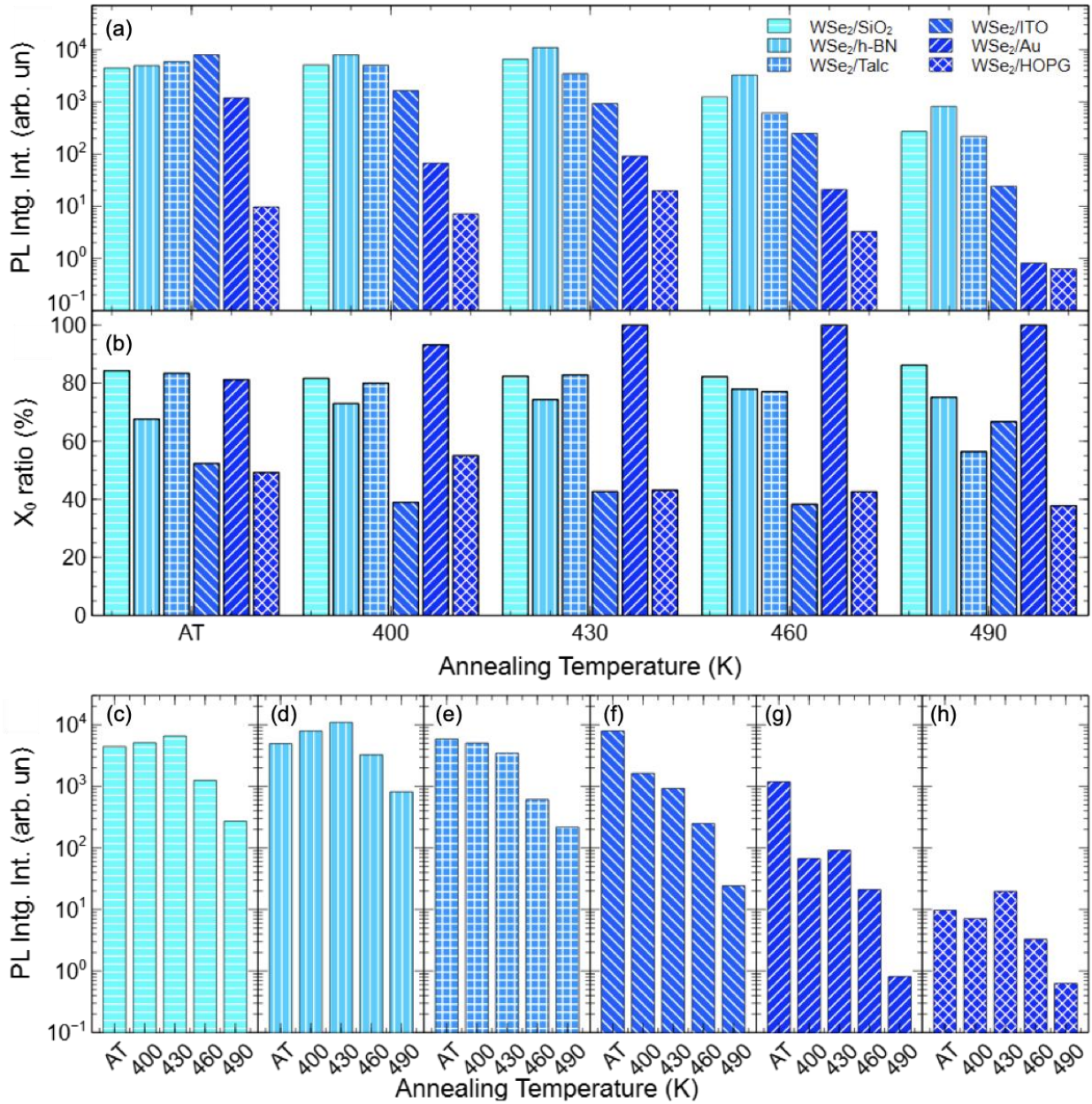

Figure S3: Evolution of the PL intensity (a) and $X_0$ contribution to the overall PL spectra (b) of ML-WSe$_2$ deposited onto several substrates, as a function of annealing temperatures. Panels (c-h) present the PL intensity measured after each annealing cycle for ML-WSe$_2$ deposited onto SiO$_2$ (c), h-BN (d), Talc (e), ITO (f), Au (g) and HOPG (h).

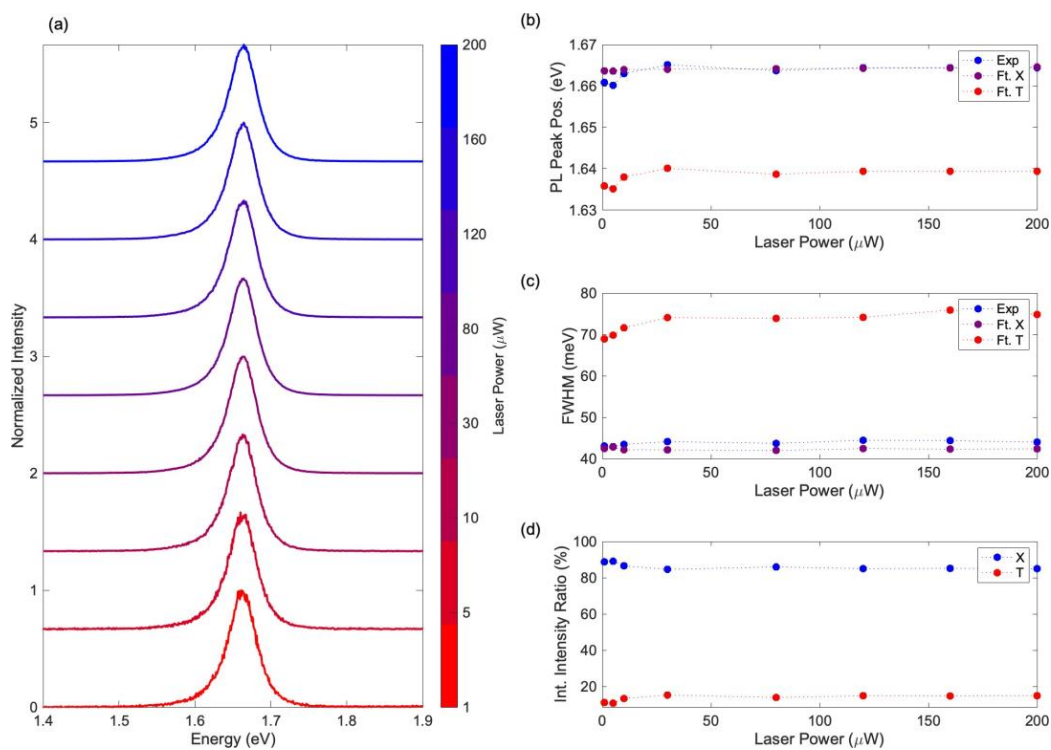

Figure S4: Influence of the laser excitation power on the luminescence of ML-WSe$_2$/SiO$_2$ after an annealing at 400 K. (a) Stacked series of normalised PL spectra as a function of the laser power. (b-d) Features extracted from the spectra presented in (a).

The influence of varying the power of the laser responsible for exciting the PL emissions measured is described in Fig. S4. Samples were excited with powers between 1-200 µW, and no relevant changes in spectra were observed. As an example, Fig. S4 summarises such analysis, carried out for the ML-WSe$_2$/SiO$_2$ after the first annealing cycle (400 K). Plot (a) exhibits a series of very similar normalised spectra, acquired over the aforementioned power range.

Plots (b-d) present estimations of the peak positions, full width at half maximum (FWHM) and relative PL intensity contribution extracted from the experimental data. These quantities were automatically evaluated from each spectrum by the code developed for performing the analysis of the acquired set of spectra. The data labelled as Exp in these plots represents a rough estimate of these features, based on a simple single peak model. As the spectra acquired at lower excitation power present an inferior signal-to-noise ratio, the automatic determination of these parameters may present some fluctuations. Generally, for excitation powers above 30 $\mu$W (with an increase in the signal-to-noise ratio), all these extracted quantities present a very steady tendency. Therefore, our interpretation is that within the power range used in the experiments, the excitation power does not exert a significant impact on the spectral features observed.